\documentclass[%
 reprint,
 superscriptaddress,
 showpacs,preprintnumbers,
 amsmath,amssymb,
 aps,
 prl,
 longbibliography,
]{revtex4-1}

\usepackage{graphicx}
\usepackage{dcolumn}
\usepackage{bm}
\usepackage{color}
\usepackage{ulem}

\begin{document}

\preprint{APS/123-QED}

\title{
Spontaneous Spatial Inversion Symmetry Breaking and Spin Hall Effect
in a Spin-ice Double-exchange Model on a Pyrochlore Lattice
}

\author{Hiroaki Ishizuka}
\affiliation{
Department of Applied Physics, University of Tokyo, Hongo, 7-3-1, Bunkyo, Tokyo 113-8656, Japan
}

\author{Yukitoshi Motome}%
\affiliation{
Department of Applied Physics, University of Tokyo, Hongo, 7-3-1, Bunkyo, Tokyo 113-8656, Japan
}

\date{\today}

\begin{abstract}
A formation of tetrahedral spin clusters is discovered by Monte Carlo simulation for a spin-ice type double-exchange model on a pyrochlore lattice. 
The spin-cluster phase is magnetically disordered, but breaks spatial inversion symmetry spontaneously by developing noncoplanar four-spin molecules periodically on the pyrochlore lattice. 
We find that the system exhibits a nonzero spin Hall conductivity in the spin-cluster phase. 
The result suggests that an intersite-multipole order induces the unconventional spin Hall state without the spin-orbit interaction.
\end{abstract}

\maketitle

Missing spatial inversion symmetry (SIS) influences the nature of condensed matters in a profound way. 
In particular, spontaneous breaking of SIS leads to fascinating cooperative phenomena.
A typical example is the ferroelectricity. 
Spontaneous breaking of SIS by lattice distortions in insulating materials gives rise to a macroscopic electric polarization. 
When time reversal symmetry (TRS) is broken in addition to the SIS breaking, interesting interplay appears between electric and magnetic degrees of freedom, called the multiferroicity.
For example, magnets with peculiar orders, such as a spiral order, exhibit the magnetoelectoric effect~\cite{Kimura2003,Katsura2005,Sergienko2006,Mostovoy2006}. 
A more complicated magnetic texture, called Skyrmion~\cite{Skyrme1962,Bogdanov1989,Muhlbauer2009,Muzner2010,Yu2010}, has also recently attracted much attention. 
In these systems, the relativistic spin-orbit interaction (SOI) plays an important role in connecting the SIS breaking and magnetism. 
When the system is conducting, such unusual spin textures significantly affect the transport properties through the spin Berry phase mechanism~\cite{Loss1992,Ye1999,Ohgushi2000}, which opens a possibility of applications to spintronics.
Thus, the search for spontaneous breaking of SIS, particularly in metallic systems, is a promising way to find new electromagnetic and transport phenomena in condensed matter physics.

In this Rapid Communication, we theoretically explore a new type of spontaneous breaking of SIS in a conductive system driven by the coupling between itinerant electrons and localized spins. 
For this purpose, we investigate a double-exchange (DE) model on a 3D pyrochlore lattice with spin-ice type localized moments, which is regarded as a fundamental model for metallic pyrochlore oxides.
We find that the model exhibits an interesting thermally-induced phase.
In this phase, the spins are thermally fluctuating and disordered, but form tetrahedral four-spin clusters arranged periodically on the lattice; the spin-cluster formation violates SIS without breaking TRS.
Long-range effective magnetic interactions driven by the spin-charge coupling play a role in stabilizing the peculiar SIS-broken state.
We also show that the thermally-induced spin-cluster phase exhibits the spin Hall effect (SHE) via the fluctuating noncoplanar spin textures.
This SHE is unconventional because our model does not include SOI which is a requisite for the conventional SHE~\cite{Dyakonov1971,Dyakonov1971b,Hirsch1999,Murakami2003,Sinova2004,Murakami2005,Schliemann2005}.
The result indicates that the spontaneous formation of noncoplanar spin objects without magnetic ordering, which are interpreted as intersite multipoles, can be a source of unconventional transport phenomena.

\begin{figure}
   \begin{center}
   \includegraphics[width=\linewidth]{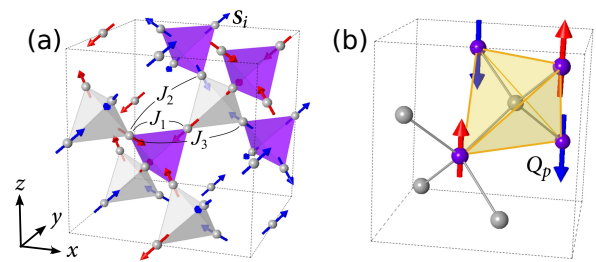} 
   \caption{
   (color online). Schematic pictures of (a) a pyrochlore lattice and (b) diamond lattice composed of the centers of tetrahedra in (a).
   In (a), the interactions in the effective Ising model in Eq.~(\ref{eq:ising}) are also shown. 
   The dark tetrahedra show the four-spin clusters with all-in or all-out configurations in the SIS-broken phase.
   In (b), the tetrahedron shows emergent frustration between the pseudospins.
   }
   \label{fig:model}
   \end{center}
\end{figure}

We here consider a DE model~\cite{Zener1951} in which itinerant electrons interact with Ising moments with spin-ice type anisotropy through the ferromagnetic Hund's-rule coupling on a pyrochlore lattice [see Fig.~\ref{fig:model}(a)]. 
We particularly consider the limit of strong Hund's-rule coupling~\cite{Anderson1955}. 
The Hamiltonian is given as 
\begin{eqnarray}
H = - \sum_{\langle i,j \rangle} ( t_{ij} c^\dagger_{i} c_{j} + \text{H.c.} ) + J_\mathrm{AFM}\sum_{\langle i, j \rangle} {\bf S}_i\cdot {\bf S}_j.
\label{eq:H}
\end{eqnarray}
Here, $c_i$ ($c_i^\dagger$) is the annihilation (creation) operator of an itinerant electron at the $i$th site, whose effective transfer integral $t_{ij}$ depends on the relative angle of neighboring Ising spins, given by $t_{ij} = t(\cos\frac{\theta_i}{2}\cos\frac{\theta_j}{2}+ \sin\frac{\theta_i}{2}\sin\frac{\theta_j}{2}e^{-{\rm i}(\varphi_i-\varphi_j)})$, where $\mathbf{S}_i=(\sin\theta_i \cos\varphi_i, \sin\theta_i \sin\varphi_i, \cos\theta_i)$.
The anisotropy axis of the Ising spin is site-dependent and along the local $[111]$ direction, as shown in Fig.~\ref{fig:model}(a).
The second term in Eq.~(\ref{eq:H}) is the antiferromagnetic (AFM) interaction between the Ising spins. 
The sum $\langle i,j \rangle$ is taken over nearest-neighbor (NN) sites on the pyrochlore lattice. 
This is a minimal model including the $[111]$ anisotropy, spin-charge coupling, and geometrical frustration, which are all present in many pyrochlore oxides~\cite{Gardner2010}. 
Hereafter, we set the energy unit $t=1$, the lattice constant of cubic unit cell $a = 1$, the Boltzmann constant $k_{\rm B} = 1$, and the unit of conductance $e/2\pi = 1$ ($e$ is the elementary charge).

In the following, we focus on the competition between different electronic and magnetic phases in the model in Eq.~(\ref{eq:H}) at quarter filling of electrons, $n=\frac{1}{N} \sum_i \langle c_i^\dagger c_i \rangle=1/4$ ($N$ is the number of sites). 
Similar problems were studied for the case with Heisenberg localized moments~\cite{Motome2010a,Motome2010b}.
In the present Ising case, the FM DE interaction favors a two-in two-out configuration of Ising spins in each tetrahedron, while the AFM interaction $J_\mathrm{AFM}$ prefers all-in or all-out.
Itinerant electrons mediate complicated interactions, which lead to a much more interesting phase competition compared with the spin-ice problem~\cite{Melko2004}.

Before going into the direct simulation of the model in Eq.~(\ref{eq:H}), which is highly cpu demanding, we first try to capture the overall picture of the phase competition by analyzing an effective spin model with kinetic-driven interactions.
To derive the effective spin model, we consider a perturbation expansion for the hopping term in Eq.~(\ref{eq:H}). 
Considering only the amplitude of $t_{ij}$ for simplicity, we rewrite it into $|t_{ij}| = t_{ij}^+ + t_{ij}^- \tilde{S}_i \tilde{S}_j $, where $t_{ij}^\pm = \frac1{2\sqrt2} (\sqrt{1+{\bf n}_i\cdot{\bf n}_j} \pm \sqrt{1-{\bf n}_i\cdot{\bf n}_j})$.
Here, $\tilde{S}_i=\pm1$ is the projected spin parameter to a local $[111]$ vector ${\bf n}_i$; ${\bf S}_i = \tilde{S}_i{\bf n}_i$.
We define ${\bf n}_i$ so that the all-in/all-out order in terms of ${\bf S}_i$ (long-range order of alternating all-in and all-out tetrahedra) corresponds to the FM order in terms of $\tilde{S}_i$.
Then, we perform a perturbation calculation up to the second order by treating the $t_{ij}^{-}$ term as the perturbation to the $t_{ij}^+$ term.
Replacing the electronic part by the unperturbed Green's functions, we end up with the effective Ising model with long-range and multiple-spin interactions.

Among many contributions, for simplicity, we consider only two-spin interactions~\cite{note_4spin};
\begin{eqnarray}
H_\mathrm{eff} &=& -J_1\sum_{\langle i,j \rangle} \tilde{S}_i\tilde{S}_j + J_2\sum_{\left\{ i, j \right\}} \tilde{S}_i\tilde{S}_j + J_3\sum_{\left[ i, j \right]} \tilde{S}_i\tilde{S}_j. \label{eq:ising}
\end{eqnarray}
Here, the estimates of the perturbation for NN, second-neighbor, and third-neighbor couplings gives $J_1=-4.19161\times 10^{-2} + J_\mathrm{AFM}/3$, $J_2=9.65132\times 10^{-4}$, and $J_3=9.96332\times 10^{-4}$, respectively [Fig.~\ref{fig:model}(a)].
Note that $J_1$ consists of two contributions: the FM DE interaction and AFM interaction $J_\mathrm{AFM}$ (the signs are reversed due to the projection from ${\bf S}_i$ to $\tilde{S}_i$).

We investigate the phase diagram of the model in Eq.~(\ref{eq:ising}) by a classical Monte Carlo (MC) simulation while varying $J_1$~\cite{note_phasediagram}.
For efficient MC sampling, we adopt, in addition to the single-spin update, a tetrahedron update, in which four spins in a tetrahedron are flipped at once, by using the heat bath method.
The calculations were typically done with $1.2\times 10^6$ ($4.9\times 10^6$) MC steps for $N=4\times 6^3$ and $4\times 8^3$ ($N=4\times 10^3$ and $4\times 12^3$) after the thermalization of $2.2\times 10^5$ ($9.2\times 10^5$) MC steps.

\begin{figure}
   \begin{center}
   \includegraphics[width=\linewidth]{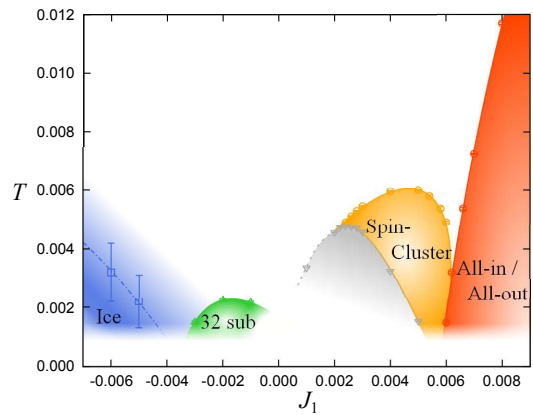} 
   \caption{
   (color online).
   Finite $T$ phase diagram for the effective Ising model in Eq.~(\ref{eq:ising}).
   The symbols indicate the critical temperatures (crossovers in the ice region) obtained by MC simulation, and the lines are the guides for eyes.
   }
   \label{fig:diagram}
   \end{center}
\end{figure}

Figure~\ref{fig:diagram} shows the phase diagram obtained by the MC simulation. 
We identify four dominant regions at low $T$: (i) the ice state for $J_1 \lesssim -0.004$, (ii) 32-sublattice ordered phase for $-0.003 \lesssim J_1 \lesssim 0$, (iii) spin-cluster phase with SIS breaking for $0.002\lesssim J_1 \lesssim 0.006$, and (iv) all-in/all-out ordered phase for $J_1\gtrsim 0.006$.

\begin{figure}
   \begin{center}
   \includegraphics[width=\linewidth]{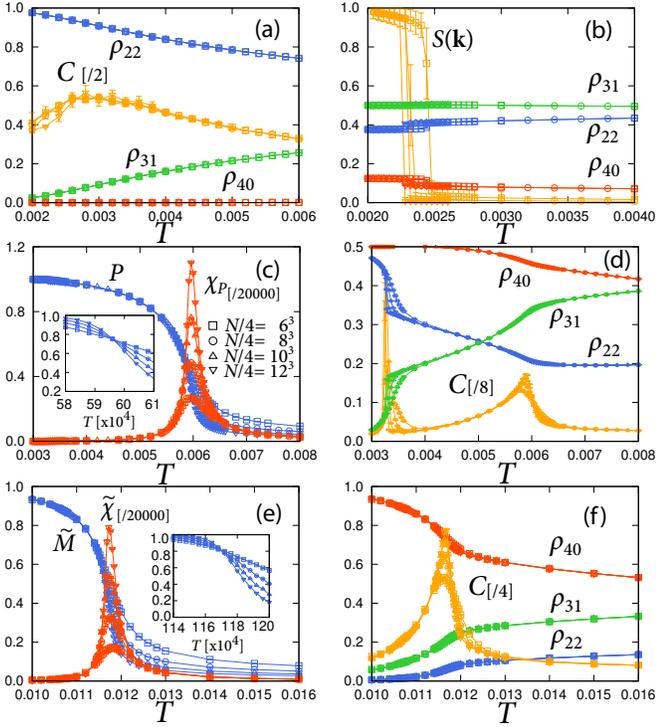} 
   \caption{
   (color online).
   MC results for the model in Eq.~(\ref{eq:ising}) at (a) $J_1=-0.006$, (b) $J_1=-0.002$, (c), (d) $J_1=0.004$, and (e), (f) $J_1=0.008$. See the text for details.
   }
   \label{fig:clmc}
   \end{center}
\end{figure}

Figure~\ref{fig:clmc} shows typical MC data for the $T$ dependence of physical quantities used for identifying these four regions.
In the region (i) where the FM DE interaction is dominant, a crossover to the two-in two-out ice state is observed in a hump of the specific heat $C$ accompanied by an increase of the fraction of two-in two-out tetrahedra, $\rho_{22}$, as shown in Fig.~\ref{fig:clmc}(a).
The situation is similar to that in the spin ice~\cite{Harris1997,Ramirez1999}.
On the other hand, in the region (iv), $J_\mathrm{AFM}$ dominates the FM DE interaction and stabilizes the all-in/all-out order. 
The transition is characterized by a rapid increase of the net magnetic moment for projected spins, $\tilde{M} = \frac{1}{N} |\sum_i \langle \tilde{S}_i \rangle|$, divergence of its susceptibility $\tilde{\chi}$ [Fig.~\ref{fig:clmc}(e)], a sharp peak in $C$, and growth of all-in and all-out fraction, $\rho_{40}$ [Fig.~\ref{fig:clmc}(f)].
The critical temperature is estimated at $T_c=0.01170(4)$ at $J_1=0.008$ from the Binder analysis of $\tilde{M}$ shown in the inset of Fig.~\ref{fig:clmc}(e)~\cite{Binder1981}.

Between the two regimes, we found interesting phases resulting from competition between the DE interaction and $J_\mathrm{AFM}$. 
One is the 32-sublattice ordered phase in the region (ii) next to the ice state, which is characterized by an abrupt increase of the spin structure factor for the same sublattice, $S({\bf k})$ at ${\bf k}=(\pi,\pi,\pi)$ [Fig.~\ref{fig:clmc}(b)].
The same ordering was recently reported in a similar model with a relatively weak Hund's-rule coupling~\cite{Ishizuka2012,note_32sub}. 

A more interesting finding here is the spin-cluster phase in the region (iii) on the verge of the all-in/all-out order. 
The transition is characterized by $P$ parameter and its susceptibility $\chi_P$ [Fig.~\ref{fig:clmc}(c)] as well as the specific heat [Fig.~\ref{fig:clmc}(d)]; here, $P$ is defined by the difference of fractions of all-in/all-out tetrahedra between upward and downward tetrahedra (two different tetrahedra in the four-site unit cell in the pyrochlore lattice), $P=|\rho_{40}^{\uparrow}-\rho_{40}^{\downarrow}|$.
The result indicates that the upward and downward tetrahedra become inequivalent at low $T$; one of them has larger population of the all-in/all-out tetrahedra than the other [see Fig.~\ref{fig:model}(a)]. 
In other words, four-spin clusters are formed and arranged periodically (the translational symmetry is not broken as the primitive unit cell includes a pair of upward and downward tetrahedra).
The transition is continuous and the critical temperature is estimated at $T_c=0.00596(2)$ at $J_1=0.004$ from the Binder analysis in the inset.
Interestingly, the phase below $T_c$ does not show any magnetic ordering; no singularity is found in $S({\bf k})$.
Therefore, the spin-cluster phase can be viewed as a classical spin-liquid state where TRS is preserved but SIS is broken due to the differentiation of upward and downward tetrahedra.

The SIS-broken phase (iii) appears only at finite $T$, as shown in Fig.~\ref{fig:diagram}; the system exhibits another transition at a lower $T$ [see also Fig.~\ref{fig:clmc}(d)]. 
This suggests that the phase (iii) is a thermally-induced intermediate phase, which is often seen in geometrically frustrated systems.
To see the frustration effect more explicitly, let us rewrite the model in Eq.~(\ref{eq:ising}) into a pseudospin model, $H_\mathrm{eff} = J_2 \sum_{\langle p,q\rangle} Q_{p}Q_{q} -\frac{\tilde{J}_1}2 \sum_p Q_p^2 + {\rm const.}$, defined on a diamond lattice composed of the centers of tetrahedra in the pyrochlore lattice [see Fig.~\ref{fig:model}(b)].
Here, $\tilde{J}_1=J_1/2+J_2$ (we take $J_2=J_3$ for simplicity) and $Q_p=\sum_{i\in p} \tilde{S}_i$ is a pseudospin at $p$th site on the diamond lattice, defined by the sum of four projected spins belonging to $p$th tetrahedron in the pyrochlore lattice~\cite{note_pseudospin}. 
This pseudospin picture maps the SIS-broken spin-cluster state to the system of the pseudomoments with $Q_p=\pm 4$ bridged by $Q_p=0$ on the diamond lattice, as shown in Fig.~\ref{fig:model}(b). 
This is effectively an Ising model on a face-centered-cubic (FCC) lattice. 
The system, therefore, hinders severe frustration in the superlattice of tetrahedra, which presumably leads to the emergence of the peculiar intermediate phase (iii)~\cite{note_on_fcc}.

\begin{figure}
   \begin{center}
   \includegraphics[width=\linewidth]{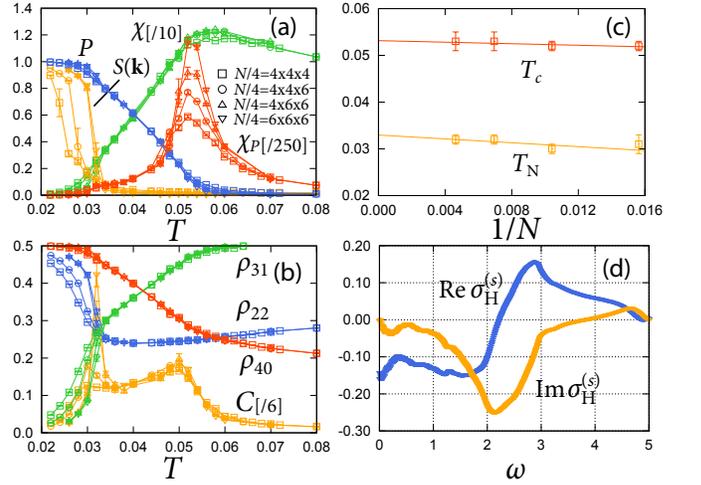}
   \caption{
   (color online).
   (a), (b) MC results for the DE model in Eq.~(\ref{eq:H}) at $n=1/4$ and $J_\mathrm{AFM}=0.18$.
   (c) System-size extrapolation of critical temperatures estimated by the peaks of $\chi_P$ and $C$.
   (d) Spin Hall conductivity calculated by taking simple average in the SIS-broken spin-cluster manifold.
   }
   \label{fig:pemc}
   \end{center}
\end{figure}

Now, we examine whether such a peculiar SIS-broken phase appears in the original DE model in Eq.~(\ref{eq:H}).
For this purpose, we here conducted the direct MC simulation of the model by using the polynomial expansion method~\cite{Motome1999,Ishizuka2013}.
We used 34 polynomials for sufficient convergence.
The calculations were done by single-spin flip and tetrahedron updates for typically 2900 MC steps after 700 MC steps of initial relaxation.

Figures~\ref{fig:pemc}(a) and \ref{fig:pemc}(b) show the MC results at $J_\mathrm{AFM}=0.18$.
The increase of $P$ and peaks of $\chi_P$ and $C$ at $T \sim 0.053$, along with the absence of anomaly in $S({\bf k})$ indicate the emergence of the SIS-broken spin-cluster phase, similar to that in Figs.~\ref{fig:clmc}(c) and \ref{fig:clmc}(d).
On the other hand, $S({\bf k})$ at ${\bf k}=(0,0,2\pi)$ sharply increases at a lower $T$ associated with a sharp peak in $C$.
The extrapolation with respect to $1/N$ of these transition temperatures (peaks of $\chi_P$ and $C$) is presented in Fig.~\ref{fig:pemc}(c).
The results show that the model in Eq.~(\ref{eq:H}) exhibits two transitions at different $T$: one is the peculiar transition which breaks only SIS by the differentiation of upward and downward tetrahedra at $T_c=0.053(2)$, and the other is the magnetic transition with additional TRS breaking at $T_{\mathrm{N}}=0.033(1)$~\cite{note_edos}.
We note that the system remains metallic below $T_c$ while it becomes insulating below $T_\mathrm{N}$.

It is also worthy noting that the magnetic susceptibility $\chi$ shows a steep decrease in the SIS-broken phase, as shown in Fig.~\ref{fig:pemc}(a).
This is in sharp contrast to the diverging Curie-law like behavior in the classical spin-liquid state in spin ice~\cite{Jaubert2013}.
The decrease of $\chi$ is likely to come from the formation of all-in/all-out clusters, in which four spins are coupled antiferromagnetically.

Interestingly, the intermediate metallic phase shows a nonzero spin Hall conductivity.
Figure~\ref{fig:pemc}(d) shows the result of spin Hall conductivity $\sigma_{\mathrm{H}}^{(\mathrm{s})}$ calculated by the Kubo formula with scattering rate of $\tau^{-1}/t=0.01$~\cite{note_js} for $N_{\bf k}=8^3$ sites of $N=4\times 4^3$ site supercells. 
As MC simulation did not reach enough convergence because of the small system sizes, we here calculated $\sigma_{\mathrm{H}}^{(\mathrm{s})}$ by assuming an ideal situation, i.e., by taking simple average over $128$ different spin configurations randomly generated so that the upward tetrahedra are either all-in or all-out.
In the calculation, an electronic field is applied along the $[110]$ direction, and the spin current and magnetic moment are measured along $[\bar{1}12]$ and $[1\bar{1}1]$, respectively.
As shown in Fig.~\ref{fig:pemc}(d), the real part of $\sigma_{\mathrm{H}}^{(\mathrm{s})}$ remains nonzero in the static limit of $\omega\to 0$.  
This suggests that the intermediate SIS-broken phase is indeed a spin Hall state.
The present SHE is a consequence of characteristic noncoplanar spin textures~\cite{Loss1992,Ye1999,Ohgushi2000}, which is distinct from the conventional SHE originating from the relativistic SOI.

The SIS breaking in our model in Eq.~(\ref{eq:H}) takes place by formation of four-spin clusters. 
Cluster formation is a manifestation of competing interactions in frustrated itinerant electron systems~\cite{Shimomura2004,Shimomura2005}. 
Our SIS-broken phase, however, retains neither charge ordering nor magnetic dipole ordering, suggesting that it is characterized by higher-order electric and magnetic multipoles. 
Hence, our results indicate that the scattering of electrons by such multipoles can lead to unconventional transport phenomena. 
Multipole orders, which are often called ``hidden orders", have attracted interests not only in localized spin systems but also in conducting systems~\cite{Mydosh2011}.
SHE may provide a further insight into such hidden multipoles.

As mentioned above, our model is simple but includes some essential features in metallic pyrochlore oxides, which have recently attracted growing interest both experimentally and theoretically~\cite{Gardner2010}. 
It is intriguing that some pyrochlore compounds indeed exhibit similar SIS breaking accompanied by a breathing-type lattice distortion~\cite{Yamaura2002,Ohgushi2011}. 
Our result suggests a possibility to observe the unconventional SHE in such class of materials.

Conventional SHE relies on the strong SOI and device structure which are both hard to control once the system is fabricated.
Our proposal, on the other hand, does not need SOI and is solely based on the competing magnetic interactions which are controllable by external stimuli, such as an applied magnetic field.
Such controllability might be useful not only for potential applications to electronic devices but also for experimentally distinguishing the origin of SHE.
Furthermore, our new SHE without SOI suggests a new direction of searching SHE materials, which might be beneficial for industrial applications.

The authors thank C. D. Batista, N. Furukawa, H. Kusunose, K. Ohgushi, K. Penc, N. Shannon, and A. Shitade for fruitful discussions.
Part of the calculations were performed on the Supercomputer Center, Insitute for Solid State Physics, University of Tokyo.
H.I. is supported by Grant-in-Aid for JSPS Fellows.
This research was supported by KAKENHI (No.~19052008, 21340090, 22540372, and 24340076), Global COE Program ``the Physical Sciences Frontier", the Strategic Programs for Innovative Research (SPIRE), MEXT, and the Computational Materials Science Initiative (CMSI), Japan.


\begin{thebibliography}{99}
\bibitem{Kimura2003}        T. Kimura, T. Goto, H. Shintani, K. Ishizaka, T. Arima, Y. Tokura, Nature {\bf 55}, 426 (2003).
\bibitem{Katsura2005}       H. Katsura, N. Nagaosa, and A. V. Balatsky, Phys. Rev. Lett. {\bf 95}, 057205 (2005). 
\bibitem{Sergienko2006}     I. A. Sergienko and E. Dagotto, Phys. Rev. B {\bf 73}, 094434 (2006). 
\bibitem{Mostovoy2006}      M. Mostovoy, Phys. Rev. Lett. {\bf 96}, 067601 (2006).
\bibitem{Skyrme1962}        T. H. R. Skyrme, Nucl. Phys. {\bf 31}, 556 (1962).
\bibitem{Bogdanov1989}      A. N. Bogdanov and D. A. Yablonskii, Sov. Phys. JETP {\bf 68}, 101 (1989).
\bibitem{Muhlbauer2009}	    S. M\"{u}hlbauer, B. Binz, F. Jonietz, C. Pfleiderer, A. Rosch, A. Neubauer, R. Georgii, P. B\"{o}ni, Science {\bf 323}, 915 (2009).
\bibitem{Muzner2010}	    W. M\"{u}nzer, A. Neubauer, T. Adams, S. M\"{u}hlbauer, C. Franz, F. Jonietz, R. Georgii, P. B\"{o}ni, B. Pedersen, M. Schmidt, A. Rosch, and C. Pfleiderer, Phys. Rev. B {\bf 81}, 041203R (2010).
\bibitem{Yu2010}            X. Z. Yu, Y. Onose, N. Kanazawa, J. H. Park, J. H. Han, Y. Matsui, N. Nagaosa, and Y. Tokura, Nature {\bf 465}, 901 (2010). 
\bibitem{Loss1992}          D. Loss and P. M. Goldbart, Phys. Rev. B {\bf 45}, 13544 (1992).
\bibitem{Ye1999}            J. Ye, Y. B. Kim, A. J. Millis, B. I. Shraiman, P. Majumdar, and Z. Te$\rm\check{s}$anovic, Phys. Rev. Lett. {\bf 83}, 3737 (1999).
\bibitem{Ohgushi2000}       K. Ohgushi, S. Murakami, and N. Nagaosa, Phys. Rev. B {\bf 62}, 6065(R) (2000).
\bibitem{Sinova2004}        J. Sinova, D. Culcer, Q. Niu, N. A. Sinitsyn, T. Jungwirth, and A. H. MacDonald, Phys. Rev. Lett. {\bf 92}, 126603 (2004).
\bibitem{Murakami2005}      S. Murakami, Adv. in Solid State Phys. {\bf 45}, 192 (2005).
\bibitem{Schliemann2005}    J. Schliemann, Int. J. Mod. Phys. B, {\bf 20}, 1015 (2005).
\bibitem{Dyakonov1971}      M. I. Dyakonov and V. I. Perel, Zh. Eksp. Teor. Fiz. Pis'ma Red. {\bf 13}, 657 (1971).
\bibitem{Dyakonov1971b}	    M. I. Dyakonov and V. I. Perel, Sov. Phys. JETP Lett. {\bf 13}, 467 (1971).
\bibitem{Hirsch1999}	    J. E. Hirsch, Phys. Rev. Lett. {\bf 83}, 1834 (1999).
\bibitem{Murakami2003}      S. Murakami, N. Nagaosa, and S. C. Zhang, Science {\bf 301}, 1348 (2003).
\bibitem{Zener1951}         C. Zener, Phys. Rev. {\bf 82}, 403 (1951).
\bibitem{Anderson1955}      P. W. Anderson and H. Hasegawa, Phys. Rev. {\bf 100}, 675 (1955).
\bibitem{Gardner2010}       J. S. Gardner, M. J. P. Gingras, and J. E. Greedan, Rev. Mod. Phys. {\bf 82}, 53 (2010).
\bibitem{Motome2010a}	    Y. Motome and N. Furukawa, Phys. Rev. Lett. {\bf 104}, 106407 (2010).
\bibitem{Motome2010b}	    Y. Motome and N. Furukawa, Phys. Rev. B {\bf 82}, 060407(R) (2010).
\bibitem{Melko2004}	    R. G. Melko and M. J. P. Gingras, J. Phys.: Cond. Matter {\bf 16}, R1277 (2004). 
\bibitem{note_4spin}        We also evaluated the four-spin interactions, but they are much smaller and subdominant compared to the two-spin interactions.
\bibitem{note_phasediagram} The overall phase diagram by changing $J_1$, $J_2$, and $J_3$ will be reported elsewhere.
\bibitem{Harris1997}        M. J. Harris, S. T. Bramwell, D. F. McMorrow, T. Zeiske, and K. W. Godfrey, Phys. Rev. Lett. {\bf 79}, 2554 (1997).
\bibitem{Ramirez1999}       A. P. Ramirez, A. Hayashi, R. J. Cava, R. Siddharthan, and B. S. Shastry, Nature (London) {\bf 399}, 333 (1999).
\bibitem{Binder1981}        K. Binder, Z. Phys. B {\bf 43}, 119 (1981).
\bibitem{Ishizuka2012}      H. Ishizuka, M. Udagawa, and Y. Motome, J. Phys. Soc. Jpn. {\bf 81}, 113706 (2012).
\bibitem{note_32sub}        In Ref.~\cite{Ishizuka2012}, the phase transition to the 32-sublattice ordered state appears to be second order. It is, however, apparently first order in the current case. We find that the order of the transition changes with $J_2/J_3$.
\bibitem{note_pseudospin}   The indices $p$ and $q$ run both upward and downward tetrahedra; hence, there is a local constraint for neighboring $Q_p$ and $Q_{q}$, as they share $\tilde{S}_i$ in between them.
\bibitem{note_on_fcc}       The transition at $T\simeq 0.00324$ might be ascribed to a dimensional reduction similar to that in the FCC Ising model; see J. L. Lebowitz, M. K. Phani, and D. F. Styer, J. Stat. Phys. {\bf 38}, 413 (1985).
\bibitem{Motome1999}        Y. Motome and N. Furukawa, J. Phys. Soc. Jpn. {\bf 68}, 3853 (1999). 
\bibitem{Ishizuka2013}      H. Ishizuka, M. Udagawa, and Y. Motome, preprint (arXiv:1307.2942), to be published in Comp. Phys. Commun.
\bibitem{note_edos}         In the lowest-$T$ phase, the upward tetrahedra form an alternating stacking of all-in planes and all-out planes along the $z$ axis, while all downward tetrahedra become two-in two-out. The result suggests that the interactions beyond the effective model in Eq.~(\ref{eq:ising}) lift the degeneracy in the FCC pseudospin model.
\bibitem{note_js}           Importance of the vertex correction was pointed out in the Rashba model, which cancels out the Hall conductance for $\tau \ne 0$. See J.-I. Inoue, G. E. W. Bauer, and L. W. Molenkamp, Phys. Rev. B {\bf 70}, 041303(R) (2004).
\bibitem{Jaubert2013}       L. D. C. Jaubert, M. J. Harris, T. Fennell, R. G. Melko, S. T. Bramwell, and P. C. W. Holdsworth, Phys. Rev. X {\bf 3}, 011014 (2013).
\bibitem{Shimomura2004}     Y. Shimomura, S. Miyahara, and N. Furukawa, J. Phys. Soc. Jpn. {\bf 73}, 1623 (2004).
\bibitem{Shimomura2005}     Y. Shimomura, S. Miyahara, and N. Furukawa, J. Phys. Soc. Jpn. {\bf 74}, 661 (2005).
\bibitem{Mydosh2011}        J. A. Mydosh and P. M. Oppeneer, Rev. Mod. Phys. {\bf 83}, 1301 (2011).
\bibitem{Yamaura2002}	    J. Yamaura and Z. Hiroi, J. Phys. Soc. Jpn. {\bf 71}, 92598 (2002).
\bibitem{Ohgushi2011}	    K. Ohgushi, J. Yamaura, M. Ichihara, Y. Kiuchi, T. Tayama, T. Sakakibara, H. Gotou, T. Yagi, and Y. Ueda, Phys. Rev. B {\bf 83}, 125103 (2011).
\end{thebibliography}
\end{document}